# Photonic Contributions to the Apparent Seebeck Coefficient of Plasmonic Metals

*Boqin Zhao[1], Annika Lee[1], Ju Eun Yim[1], Zachary Brawley[2], Emma Brass[3], Matthew Sheldon[1,2,3,4]\**

[1]Department of Chemistry, Texas A&M University, College Station, TX 77843, USA.

[2]Department of Materials Science and Engineering, Texas A&M University, College Station, TX 77843, USA.

[3]Department of Chemistry, University of California, Irvine, Irvine, CA 92697, USA.

[4]Department of Materials Science and Engineering, University of California, Irvine, Irvine, CA 92697, United States

*Corresponding Author: Matthew Sheldon: m.sheldon@uci.edu

Photo-induced charge transport in plasmonic metal nanostructures has garnered significant interest for applications in sensing and power conversion, yet the underlying mechanisms remain debated. Here, we report spatially correlated photovoltage generation in photonically engineered Au nanowires illuminated by focused, milliwatt-level laser excitation. Plasmonic nanodisk antennas placed adjacent to the nanowires created local variations in the photonic environment, resulting in clearly defined regions of enhanced photovoltage. Experimental results and simulations strongly support a thermally driven photothermoelectric (PTE) mechanism, where the local photonic structure modifies the intrinsic Seebeck coefficient of the metal, independent of other electronic structural factors. Our findings highlight photon-electron interactions as critical to the observed transport phenomena, suggesting photonic engineering as a viable strategy to systematically control and optimize thermoelectric performance.

**Introduction**

The optical and electronic properties of metals are deeply interconnected[1,2]. The electronic structure of a metal governs its optical absorption spectrum and, more generally, its complex dielectric function. Collective charge density oscillations in the electron gas of a metal give rise to plasmon resonances, which can strongly confine optical fields and shape the corresponding photonic environment[3]. Understanding the fundamental physical relationships between these phenomena is crucial for exploring a range of applications that leverage the optoelectronic behavior of metals, for example, in photocatalysis[4–6], light detection[7], and solar energy harvesting[8].

Light-induced photovoltage and photocurrent have been observed in numerous experiments probing the optoelectronic properties of all-metal circuits defined by plasmonic nanostructures[9–21], yet there is little consensus on the fundamental mechanisms underlying this behavior. The magnitude of photovoltage is typically found to be proportional to the incident light intensity, with a characteristic scale of ~µV's under focused laser illumination at ~mW power (or ~$10^8$ W/m² power intensity). The Natelson laboratory, among others, has previously reported such photovoltage in nanoscale circuits composed of a single metal, e.g. Au, attributing the signal to photothermal heating and a resulting thermoelectric response – commonly described as a "photothermoelectric" (PTE) effect[9–15]. That is, the photovoltage is hypothesized to result from a conventional thermoelectric effect, except that temperature gradients are supplied by local photothermal heating due to focused laser excitation on the nanostructure. At every location in the circuit a thermoelectric voltage, or thermally driven electromotive force (EMF), is produced that is proportional to the local temperature gradient (Eq. 1), where the proportionality constant is the conventional Seebeck coefficient, $S$. Then, the experimentally measured, macroscopic photovoltage is interpreted as the local EMF integrated throughout the entire circuit. It is important to emphasize that a temperature gradient must occur in combination with a change in the magnitude of the Seebeck coefficient in order to generate a nonzero overall circuit voltage.

$$V = \int E_{EMF}\, dl = \int S(l) \cdot \nabla T(l)\, dl \qquad (1)$$

The magnitude of the Seebeck coefficient in a metal is traditionally understood analytically in terms of the Mott relation[14] (Eq. 2), where it is closely linked to the electronic structure near the Fermi level ($E_F$), appearing as the energy derivative of the electronic conductivity ($\sigma$) term. It has been noted that various factors such as geometrical confinement[9,10,13], crystal grain structure[14], local strain[14], surface conditions[10], and impurity doping[15] can influence the Seebeck coefficient by modifying the electronic structure and transport properties of the metal. As a result, the PTE effect can arise in a device composed of a single metal, provided that these structural electronic factors are engineered to vary spatially, thereby modifying the Seebeck coefficient. For example, it has been reported that a Au circuit junction defined by an abrupt change in width from 20 μm to 100 nm produces ~20% change in the Seebeck coefficient[13].

$$S = -\frac{\pi^2 k_B T}{3e}\frac{dln\sigma}{dE}\bigg|_{E=E_F} \qquad (2)$$

In addition to PTE effects, a variety of other photoelectrical phenomena have been described in all-metal circuits, many of which are primarily non-thermal in nature. One such mechanism is photon drag[16,17] which is understood as the transfer of photon momentum to charge carriers, resulting in directional electron transport under obliquely incident light on a thin metal film. When this effect occurs in the presence of plasmonic field enhancement, it is often referred to as "plasmon drag"[18–20]. Notably, more recent studies highlight significant challenges in fully explaining the phenomenon solely in terms of photon-to-electron momentum transfer[21]. The plasmoelectric effect[22,23], by contrast, has been described as a thermodynamically driven auto-resonance phenomenon that induces charge transport to or from plasmonic nanoparticles under light illumination. Other related opto-electronic and transport effects such as hot-electron injection[24] and gap tunneling[9,25,26] phenomena have also been observed in all-metal plasmonic nanostructures. However, significant confusion remains regarding the correct assignment of mechanisms for specific experimental observations.

These non-thermal photovoltage mechanisms highlight how photons can influence electron transport in metals through both direct and indirect interactions. Surface plasmon polaritons, in particular, correspond to a strong coupling interaction between photons and electrons at the metal surface[27,28], so that photonic and electronic states mutually affect one another[29]. That is, both the electronic and photonic density of states (DOS) are modified. Given this interplay, it is reasonable to expect that plasmonic modes in nanostructured gold may also impact electron transport processes, including those contributing to the PTE effect discussed above. The present study seeks to further clarify how photonic structure influences electronic transport phenomena in all-metal plasmonic circuits.

Here, we report the observation of photovoltage generated by Au nanowires with engineered photonic structures along their length. The observed electrical signals are unambiguously correlated with variations in the local photonic environment. While the signals share some characteristics with previously reported PTE effects, such as those by Natelson[10,12–15] and others[11], our device design deliberately eliminates structural factors known to influence the Seebeck coefficient in conventional PTE scenarios. Specifically, the nanowire devices in this study (Figure 1a) exhibit minimal variation in diameter, surface morphology, crystal strain, or related structural features. Instead, our results highlight how spatial variations in photonic structure appear to modulate electron transport properties. Although our observations may still be interpreted within the broader framework of the PTE effect, we hypothesize that the engineered photonic landscape modifies the intrinsic electronic transport in the nanowire.

**Device design**

Please see the Methods section for a full description of the device geometry. Each device consisted of an Au nanowire electrically connected to an external circuit (Figure 1a). For selected regions along the nanowire, arrays of plasmonic Au nanodisks were positioned in close proximity (~200 nm) to provide localized electromagnetic field enhancement. This design creates a sharp interface along the length of the nanowire, with distinct photonic environments on either side of the nanodisk block boundary (referred to as "in-disk" or "off-disk" throughout this text). Importantly, the nanowire itself remains structurally homogeneous, with minimal variations in width, surface chemistry, strain, or other factors previously associated with Seebeck coefficient modulation and PTE voltage generation[9–15].

We performed full-wave electromagnetic simulations using the finite-difference time-domain (FDTD) method to gain insight into the electromagnetic field enhancement behavior. As shown in Figure 1b, c, there is significant electromagnetic field concentration on the nanowire surface for infrared wavelengths (near $\lambda=10\mu m$) in the "in-disk" regions. This enhancement corresponds to the spectral band of maximum black-body radiation emission, as described by Planck's law (SI Section 1 and Figure S1), when the device is at room temperature or heated by ≤10 K due to laser excitation. In contrast, at the laser excitation wavelength of 532 nm, the optical field shows little enhancement, despite the adjacent nanodisks (Figure 1d, e). This observed spectral dependence serves as an initial indication that the transport effects reported here may be thermally driven. Additional evidence supporting this interpretation is presented below.

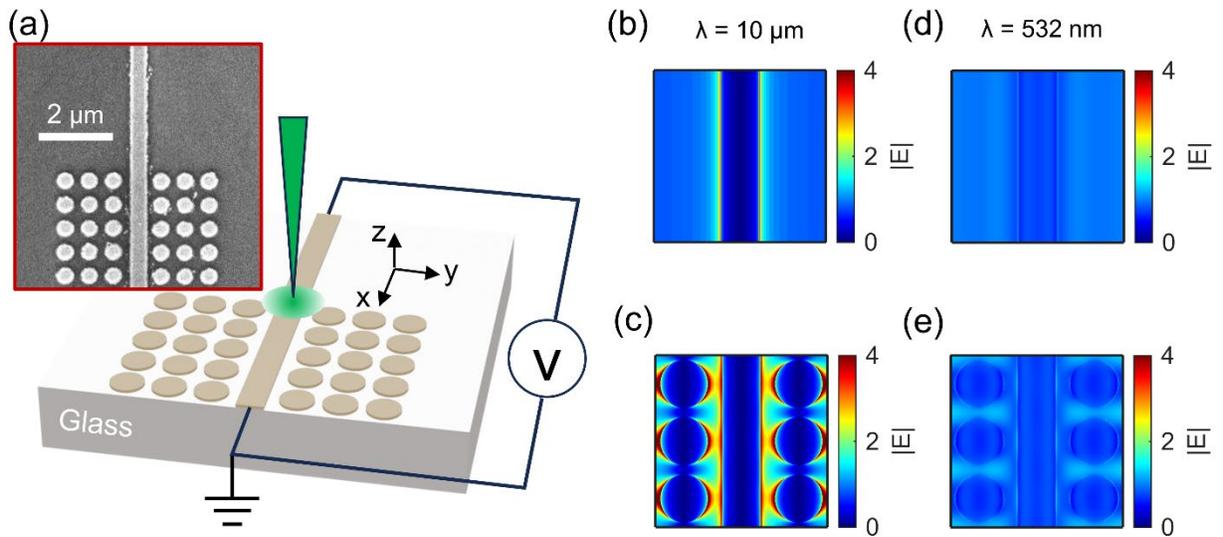

**Figure 1.** Overview of the device design, with 500 nm nanodisk diameter and 600 nm periodicity. (a) A schematic representation of the device design and the photovoltage measurement setup. Inset: scanning electron microscope (SEM) image of a test device. The nanowire and the nanodisks have a thickness of 50 nm. (b)(c) Finite difference time domain (FDTD) simulation of the electric field distribution at 10 μm wavelength, for an "off-disk" and "in-disk" region respectively under plane wave excitation. The field distribution corresponds to the plane that intersects the top surface of the nanowire. (d)(e) Similar electric field distribution at 532 nm laser excitation wavelength. The gap size between the nanowire and the nearest nanodisks is 150 nm in these simulations.

**Results and Discussions**

To characterize the optoelectronic response of the device, we raster-scanned a focused 532 nm continuous-wave (CW) laser beam (∼2 μm $1/e^2$ radius) across the nanowire and recorded the open-circuit voltage at each laser beam position, generating a two-dimensional photovoltage map. We observed clear, spatially correlated photovoltage generation, with the largest signal occurring when the laser was focused at the interface between "in-disk" and "off-disk" regions of the nanowire. As shown in Figure 2b, the sign of the photovoltage (+/−) also reverses at opposite interfaces between "in-disk" and "off-disk" regions. This is visualized by red and blue hotspots, which represent positive (+) and negative (−) measured potentials, respectively, referenced to the grounded bottom electrode. A red hotspot corresponds to a positive voltage at the top electrode, implying that electrons flow toward the grounded bottom electrode, as schematically illustrated with arrows in Figure 2a. Consequently, we infer that photoexcitation causes electrons to flow out of the "in-disk" regions into the "off-disk" regions. Additionally, the magnitude of the measured photovoltage scales linearly with the incident laser power (SI Section 2 and Figure S2). In our experiments the incident laser power remained below 1 mW ($8\times10^7$ W/m$^2$ intensity). For consistency and ease of comparison across datasets, all photovoltage amplitudes in this manuscript are normalized to the incident laser power.

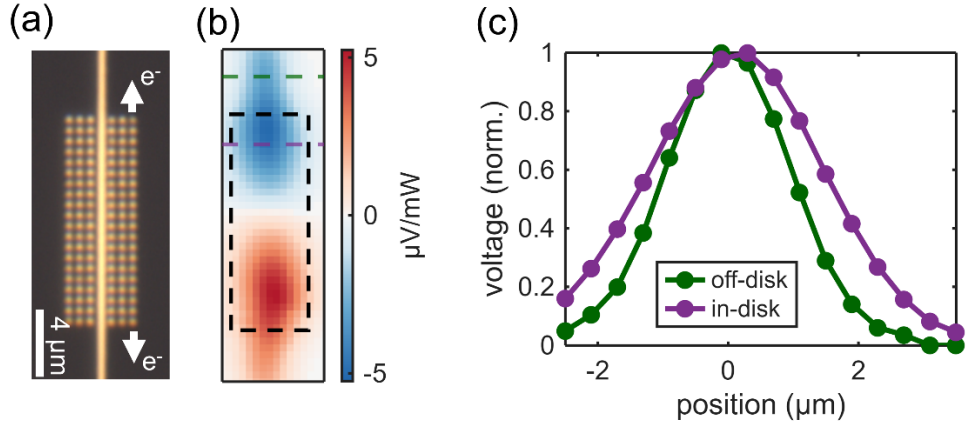

**Figure 2.** Summary of the spatially correlated photovoltage. (a) Optical bright-field image of the nanowire near a nanodisk block with ~500 nm diameter and 600 nm periodicity. The white arrows denote the direction of electron flow as indicated by the measured photovoltage signal. (b) The corresponding photovoltage map (normalized to the incident laser power) in the same region. Black dashed line denotes the location of the nanodisk block. (c) Two representative lateral line profiles from the "in-disk" region (purple trace) and "off-disk" region (green trace) corresponding to the purple and green lines in (b).

We closely examined the spatial distribution of the photovoltage map (Figure 2b). Two representative lateral line profiles from the "in-disk" region (purple trace) and "off-disk" region (green trace) are compared in Figure 2c. It is evident that the photovoltage response spans a broader lateral region in the "in-disk" region of the nanowire. To quantify this observation, we performed systematic Gaussian fittings to the horizontal slices of the 2D map (SI Section 3 and Figure S3). We found that the lateral spread of the photovoltage signal for "in-disk" regions consistently exceeds the size of the laser beam, while the spread in the "off-disk" regions is comparable to the laser beam size. This implies an energy transport mechanism from the nanodisks to the nanowire that extends beyond the direct optical excitation region—i.e., even when the laser spot does not physically overlap with the nanowire. The most plausible explanation is thermal conduction through the substrate, following photothermal heating in the nanodisks[30]. Based on this, we conclude that the 532 nm laser contributes to the observed photovoltage primarily by providing thermal energy to the system. This thermally driven behavior, combined with the fact that laser excitation in our experiments occurs at normal incidence (i.e., without lateral momentum transfer), strongly suggests that non-thermal photoelectrical mechanisms such as photon drag[21] are unlikely to play a direct role in the observed response.

Based on these observations, we aim to gain quantitative insight into the spatially correlated photovoltage by applying the theoretical framework established for the PTE effect in nanoscale plasmonic devices[10]. While we acknowledge that more complex mechanisms may also be at play, the PTE model offers a useful starting point for interpretation. Using Eq. 1 and the experimentally observed sign of the photovoltage, the Seebeck coefficient gradient must be such that the "in-disk" region of the nanowire exhibits a higher Seebeck coefficient than the "off-disk" region (i.e., $S_{off-disk} < S_{in-disk}$). Given the absence of structural differences in the nanowire between these regions, this gradient appears to arise from differences in the photonic environment. Within this interpretation, the increased photonic density of states at thermal radiation frequencies (primarily in the infrared near room temperature) in the "in-disk" region enhances the Seebeck coefficient. We estimate that laser excitation at 0.12 mW (~$1\times10^7$ W/m$^2$) used in Figure 2b produces a peak temperature rise of at most 1.5 K[31]. Based on this estimate and the magnitude of the measured photovoltage, we infer a Seebeck coefficient gradient ($\Delta S$) of at least 0.4 µV/K across the interface. This value represents 27% of the Seebeck coefficient of bulk Au (1.5 µV/K).

We also investigated the dependence of the photovoltage signal on electromagnetic field enhancement, which increases as the gap between the nanowire and adjacent nanodisks decreases. By comparing three devices with different nanowire-to-nanodisk separations—approximately 120 nm, 180 nm, and 220 nm, respectively (Figure 3a-f)—we observed that the magnitude of the spatially correlated photovoltage is inversely related to the gap size (Figure 3g-i). Devices with smaller gaps exhibited higher photovoltage signals. In contrast, for the largest gap size (~220 nm), the signal was weak and showed little to no correlation with the interfaces. This behavior is similar to that observed in control devices consisting of bare nanowires, as discussed below.

For more insight, we computed the local photonic density of states (LDOS) at a point 30 nm away from the nanowire within the gap region (Figure 3j, star symbol in inset). The simulations show that LDOS enhancement in the infrared, corresponding to the thermally active wavelengths, increases with decreasing gap size (Figure 3j), indicating stronger plasmonic coupling between the nanowire and nanodisks. In contrast, the LDOS at 532 nm remains largely unaffected by the presence of the nanodisks (SI Section 4 and Figure S4). Next, we quantitatively compared the increase in LDOS enhancement due to the nanodisks (averaged over the 10 μm to 20 μm wavelength range) with the measured photovoltage for these different devices and found a strikingly similar trend (Figure 3k). This strong correlation supports the interpretation that infrared photonic coupling effects play a significant role in generating the spatially correlated photovoltage signal.

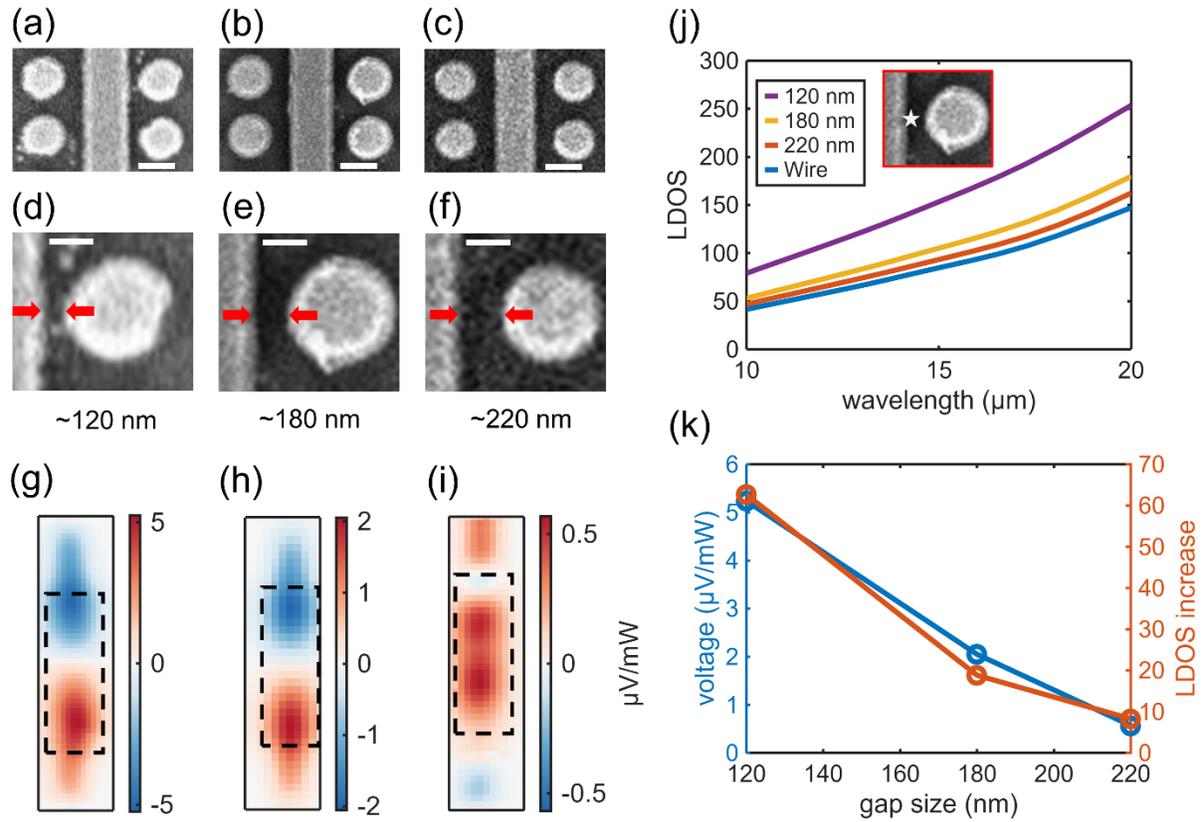

**Figure 3.** Comparison of the photovoltage map around the nanodisk block of 600 nm periodicity on three different devices with different gap sizes between the nanowire and the closest nanodisks. (a-c) Close-up SEM images of the devices. Scale bars are 400 nm. (d-f) Zoomed in SEM images at the gap region showing the gap sizes (between two red arrows) of approximately 120 nm, 180 nm and 220 nm, respectively. Scale bars are 200 nm. (g-h) Corresponding photovoltage maps (normalized to the incident laser power), with black dashed lines denoting the locations of the nanodisk block. (j) Calculated LDOS enhancement (relative to vacuum) at a position 30 nm away from the nanowire in the gap, along the top surface of the nanostructure (illustrated as the star symbol in the inset) of the three devices and a bare nanowire device. (k) LDOS enhancement increase relative to the bare nanowire averaged over the 10 μm to 20 μm wavelength range plotted versus gap size (red), compared with the maximum photovoltages versus gap size (blue).

During experimentation, we found that the magnitude of photovoltage signal increased significantly when the devices were briefly thermally annealed on a hot plate in an inert atmosphere. For the full device shown in Figure 4a, we initially observed only weak photovoltage signals immediately after nanofabrication (Figure 4b), although it was still roughly spatially correlated (Figure S5b). Annealing at 100 °C for 10 minutes did not result in a noticeable change (Figure 4c). However, the unambiguous spatially correlated photovoltage signal presented throughout this manuscript emerged after annealing at 200 °C for 10 minutes (Figure 4d), with the magnitude increasing nearly tenfold compared to the as-fabricated device (Figure 4b). In contrast, a control device lacking nanodisk antenna structures (bare nanowire device) exhibited only weak, spatially uncorrelated photovoltage signals after fabrication (Figure S6b), comparable in magnitude to the initial response of the device with nanodisks (Figure S5b). These weak signals on the bare nanowire device did not significantly change after 200 °C annealing (Figure S6c). Such spatially uncorrelated signals are consistent with previous reports attributing these weak stochastic signals to intrinsic structural defects in Au nanowires[10].

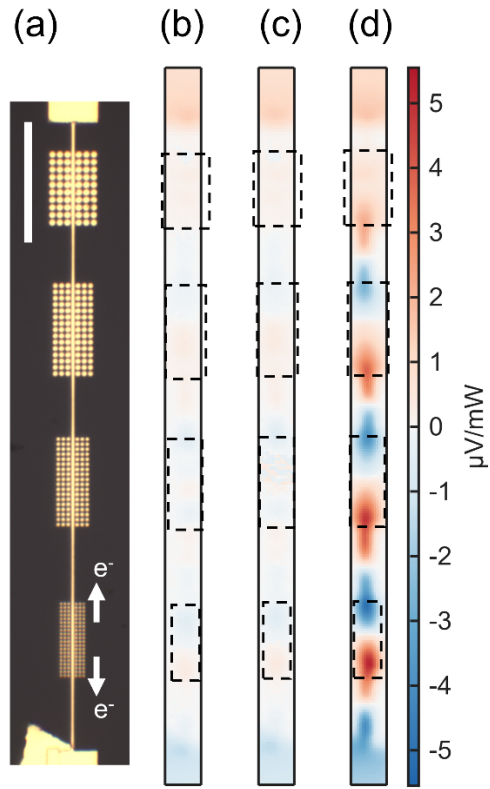

**Figure 4.** The effect of hot plate annealing on the evolution of the photovoltage. (a) Optical bright-field image of the device, where the white arrows denote the direction of electron flow as indicated by the measured photovoltage signal. The scale bar is 20 μm. (b) Photovoltage map before hot plate annealing. (c) Photovoltage map after 100 °C hot plate annealing. (d) Photovoltage map after 200 °C hot plate annealing. In (b-d), the photovoltage was normalized to the incident laser power and plotted against the same color scale. The black dashed lines denote the corresponding locations of the nanodisk blocks.

It is well established that thermal annealing of Au nanostructures reduces grain boundaries by promoting grain coalescence[32–35]. This grain coarsening improves the plasmonic field enhancement typically measured by the resonance quality factor (Q-factor)[36]. The improvement is attributed to a reduction in electron scattering at grain boundaries[36]. We therefore attribute the improved photovoltage response after annealing to enhanced plasmonic resonance quality due to reduced structural disorder. We also observed a modest decrease in device resistance (<1%) following annealing, which is also consistent with enhanced electrical conductivity across larger grains[37]. Despite these improvements, no other obvious changes were detected in the nanostructure geometry using optical microscopy (bright-field and dark-field imaging, SI Section 7 and Figure S7), suggesting that the annealing process at 200 °C does not cause melting or morphological deformation. Because thermal annealing was applied uniformly to the entire device, we expect the nanowires to be homogeneously annealed. As such, while grain structure can affect the absolute value of the Seebeck coefficient[14], it is unlikely that spatial gradients in crystallinity or grain size were introduced along the nanowire. Therefore, this uniform annealing cannot account for the observed photovoltage patterns, which require spatial variation in the Seebeck coefficient. Rather, uniform thermal annealing improved the photonic quality of the devices.

Nonetheless, we considered other non-idealities that could potentially contribute to the observed spatially correlated photovoltage signal. In particular, an unavoidable proximity effect[38] during electron-beam lithography may lead to slight broadening of the nanowire near the nanodisks. This is due to the fact that both the nanowire and nanodisks were patterned in the same resist layer, and back-scattered electrons from the electron beam exposure of the nanodisks could slightly overdose the adjacent nanowire regions, resulting in overexposed and slightly wider nanowire near those areas (see SI Section 8 for a detailed discussion). Indeed, scanning electron microscopy (SEM) images from the device reveal signs of this effect: nanowires appear slightly wider by up to 10% in the "in-disk" regions compared to the "off-disk" regions (Figure S8). Since the Seebeck coefficient of a metal is known to depend on its geometrical dimensions[13,39–42], this non-uniformity raises the concern that nanowire width variations due to proximity effects could create Seebeck coefficient gradients, and thereby contribute to the spatially correlated photovoltage.

However, this explanation is inconsistent with our experimental trends highlighted in Figure 4c. The largest photovoltage is observed at the interfaces adjacent to the smallest nanodisk block, with the smallest disk diameter and the lowest cumulative e-beam dose, where the proximity effects are expected to be minimal and the nanowire geometry is the most uniform (SI Section 8). This inverse correlation suggests that proximity-induced geometric variation is not the main mechanism. Further, a quantitative estimate based on prior experimental data[39,42] shows that even the largest observed nanowire width variations (10%) would produce Seebeck coefficient gradients far too small to explain the measured photovoltage (Analyses in SI Section 8). These findings collectively indicate that proximity effect-induced width variation is unlikely to contribute significantly.

Our experiments clearly demonstrate that the infrared photonic environment can influence the Seebeck coefficient of the nanowire, though the precise mechanism remains unclear. We propose two hypotheses. First, based on the Mott relation (Eq. 1), it is conceivable that the presence of nearby nanodisks alters the electron scattering dynamics in the nanowire—such as the damping rate or mean free path—via enhanced thermal excitation of carriers near the Fermi level. These modifications could asymmetrically perturb transport properties between the "in-disk" and "off-disk" regions. Alternatively, we consider a thermodynamic perspective on the Seebeck coefficient. Unlike the microscopic interpretation provided by the Mott relation, the Seebeck coefficient can also be viewed as the thermodynamic driving force on the electron gas to flow down a temperature gradient, which is intrinsically linked to the entropy produced per electron during this process. This entropy, in turn, depends on the electronic heat capacity of the metal. Within this framework, the local photonic environment may modify the heat capacity of the coupled electron-photon states that comprise the plasmon resonance. If so, this mechanism would resemble the previously reported plasmoelectric effect[22,23], in which charge transport arises from thermodynamic gradients acting on hybridized electron-photon states. Future studies that spatially resolve both the resistivity and temperature of the nanowire, combined with precise control over local photonic modes, could further illuminate the underlying physics at play.

**Conclusions**

In summary, we experimentally measured spatially correlated photovoltage signals in photonically engineered Au nanowires. The magnitude and sign of the photovoltage is clearly linked to local variations in the photonic environment. Our measurements confirmed that the photovoltage predominantly arises from thermally driven photothermoelectric (PTE) processes. Further, the signal was enhanced significantly by annealing-induced improvements of the plasmonic resonance quality. Detailed analysis and LDOS simulations strongly suggest that the engineered photonic structures modulate the intrinsic Seebeck coefficient of the nanowire independently of other structural variations, highlighting a key role for photon-electron interactions. This work opens possibilities for utilizing photonic degrees of freedom to systematically control and optimize thermoelectric performance, potentially informing design principles in future thermoelectric materials and devices.

**Supporting Information**

Methods, black-body radiation spectrum near room temperature, photovoltage linear laser power dependence, Gaussian fittings of the horizontal slices in the photovoltage map, LDOS enhancement in the visible spectral range, photovoltage map before and after annealing, photovoltage map on a bare nanowire device, optical images before and after annealing, possible intrinsic geometrical variations of the nanowire.

## Author contributions

B.Z. carried out the measurements, analyzed the data, performed the simulations and drafted the manuscript. A. L., J. E. Y., Z.B. and E. B. fabricated the nanostructures and performed SEM measurements. M.S. supervised the project and participated in the analysis of the data.

## Acknowledgements

This work was supported by the National Science Foundation (Grant No. CHE 2404128) and the Lockheed Martin Corporation. We would like to thank the AggieFab Nanofabrication Facility at Texas A&M University for the use of their facilities and expertise for help fabricating our nanostructures. We also thank Professor Matthew Law's lab at UC Irvine for thermal deposition of gold.

# Supplementary Information for

# Photonic Contributions to the Apparent Seebeck Coefficient of Plasmonic Metals

*Boqin Zhao[1], Annika Lee[1], Ju Eun Yim[1], Zachary Brawley[2], Emma Brass[3], Matthew Sheldon[1,2,3,4]\**

[1]Department of Chemistry, Texas A&M University, College Station, TX 77843, USA.

[2]Department of Materials Science and Engineering, Texas A&M University, College Station, TX 77843, USA.

[3]Department of Chemistry, University of California, Irvine, Irvine, CA 92697, USA.

[4]Department of Materials Science and Engineering, University of California, Irvine, CA 92697, United States

*Corresponding Author: Matthew Sheldon: m.sheldon@uci.edu

## Methods

**Device fabrication**
Each device consisted of an Au nanowire (50 nm height, 400 nm width, with a 5 nm Cr adhesion layer) with 4 nanodisk array blocks (50 nm height, with a 5 nm Cr adhesion layer) of different block sizes. These periodic nanodisk blocks from the smallest to the largest had periodicities of 600 nm, 800 nm, 1000 nm and 1200 nm, respectively. The diameter of the nanodisks in these blocks were set to leave a 200 nm gap between the adjacent nanodisks, that is, 400 nm, 600 nm, 800 nm and 1000 nm, respectively. The center-to-center distance between the nearest nanodisks and the nanowire was also targeted to leave a 200 nm gap between the nanodisks and the nanowire, that is, 600 nm, 700 nm, 800 nm and 900 nm, respectively. Glass substrates with pre-deposited Au electrodes were purchased from Platypus Tech Inc. Au nanostructure devices were fabricated on these glass substrates with electrode pads using electron-beam lithography (FEI Magellan 400 XHR SEM). PMMA resist was used to define the device geometry. A thin layer of anti-charging agent ($H_2O$ DisCharge, DisChem Inc.) was spin-coated on top of the PMMA resist layer to facilitate charge dissipation during e-beam writing. Afterwards, thermal evaporative deposition was used to deposit 5 nm Cr and 50 nm Au. Excess resist was lifted off by soaking in acetone overnight. SEM images of the devices were collected using the same SEM instrument, but only after all electrical measurements had been performed.

**Hot plate annealing**
Hot plate annealing of the devices was performed on a hot plate inside the glove box filled with Ar gas. The annealing time was at least 10 minutes at the temperatures specified in the main text.

**Photovoltage measurements**
For photovoltage measurements, the device was mounted on a confocal Raman microscope (WiTec alpha 300 RA) equipped with a piezo-electric scanning stage. Devices were electrically connected to a lock-in amplifier (Zurich Instruments HF2LI) through a voltage amplifier (Thorlabs AMP200) operating at 100V/V amplification. The top electrode was connected to the core of the BNC port of the lock-in amplifier, while the bottom electrode was connected to the shell of the BNC port. An incident 532 nm laser (CivilLaser) was modulated by a mechanical chopper (Thorlabs MC1F10HP) at 393 Hz and focused on the device with a 20x objective (NA=0.4) to a beam radius of ~2 μm ($1/e^2$ radius), maintaining a transverse polarization perpendicular to the nanowire. During the measurement, the laser was raster scanned point-by-point in a zig-zag pattern across the device and the voltage reading from the lock-in amplifier was extracted for every point.

All data collection and processing steps were automated in MATLAB. The resistance of the device was measured with a benchtop multimeter.

**Finite-difference time-domain simulations**
We performed three-dimensional full-wave simulations using finite-difference time-domain (FDTD) methods (Lumerical, Ansys Inc.). The geometry simulated consists of an infinitely thick $SiO_2$ substrate layer and the Au nanostructure on top with Cr sticking layer in between. The Au nanowire has a width of 400 nm, and the Au nanodisks are arranged periodically with a periodicity of 600 nm. The center-to-center distance between the nearest nanodisks and the nanowire is also 600 nm. The diameter of the nanodisks vary across different simulations to achieve desired gap size between the nearest nanodisks and the nanowire. $SiO_2$ was modelled with a non-dispersive refractive index of 1.5. The relative permittivity values for Au were obtained from the Drude-Lorentz model fitting of the experimental permittivity values of Au, and for Cr obtained from the Brendel-Bormann model fitting of the experimental values, both from Rakic et. al.[1] The Cr layer was 5 nm and the Au layer on top was 50 nm in thickness.

For simulations of the electric field distribution and the absorption cross section of the nanostructure device, we used the plane wave excitation source. The plane wave source was illuminating from above the device, propagating in the -z direction (as in Figure 1a) along the normal direction to the substrate surface. In these simulations, the gap size of the device was set to 150 nm.

We also evaluated the local photonic density of states (LDOS) near the Au nanowire. We simulated multiple device geometries with the gap sizes of 120 nm, 180 nm and 220 nm, respectively. We also simulated the nanowire without the nanodisks. We calculated the relative LDOS (that is, the LDOS on the device relative to the vacuum LDOS) $\rho_{rel}$ at a position 30 nm away from the edge of the nanowire at the same height as the top surface of Au. In the "in-disk" regions of the nanowire, we probed the position directly in the gap between the nanowire and the nanodisk, also 30 nm away from the edge of the nanowire at the same height as the top surface of Au. In the FDTD simulation, a test dipole source with a dipole orientation $i$ (along x, y or z direction, $i = x, y\ or\ z$) was placed at the desired position near the nanowire. The amplitude of the dipole source was fixed, and the actual power emitted by the dipole source $P_i$ was measured with a transmission box enclosing the dipole source. The relative partial local photonic density of states (PLDOS) $\rho_{rel,i}$ of a given dipole orientation was then calculated as the ratio of the actual emitted power to the radiated power $P_0$ from the same dipole source in vacuum[2], $\rho_{rel,i} = P_i/P_0$. Later, the relative LDOS $\rho_{rel}$ was achieved by averaging over all polarization directions, $\rho_{rel} = (\rho_{rel,x} + \rho_{rel,y} + \rho_{rel,z})/3$.

## Additional Analyses
### 1. Black-body radiation spectrum near room temperature

According to Planck's law, any object at a temperature $T > 0\ K$ emits thermal radiation. For an ideal black-body, the spectral radiance of thermal radiation per unit surface area, per unit solid angle, and per unit frequency is:

$$B_\nu(T) = \frac{2h\nu^3}{c^2} \frac{1}{e^{\frac{h\nu}{kT}} - 1}, \qquad (S1)$$

where $h$ is Planck's constant, $c$ is the speed of light, $k$ is Boltzmann's constant, $\nu$ is the frequency of radiation, $T$ is the temperature. The total spectral intensity integrated over the hemisphere from the surface is:

$$I_\nu(T) = B_\nu(T) \cdot \pi = \frac{2\pi h\nu^3}{c^2} \frac{1}{e^{\frac{h\nu}{kT}} - 1}. \qquad (S2)$$

Based on this formula, we calculated the total spectral radiance at a temperature $T = 300\ K$, $T = 350\ K$ and $T = 400\ K$ respectively, shown in Figure S1. At these temperatures, the majority of thermal radiation lies in the infrared spectral region roughly between 500-1000 cm$^{-1}$ in wavenumber, or correspondingly 10-20 µm in wavelength.

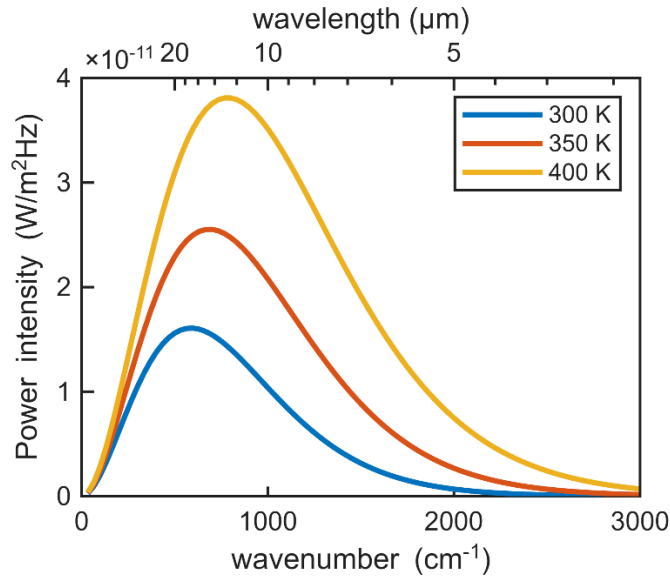

**Figure S1.** Black-body radiation spectrum per unit surface area and per unit frequency at 300 K, 350 K and 400 K, respectively.

## 2. Photovoltage linear laser power dependence

We found that the magnitude of photovoltage scales linearly with the incident laser power (Figure S2). Such linear relationship has also been observed in other related photovoltage generation mechanisms such as the photothermoelectric (PTE) effect[3] and the photon drag effect[4]. Therefore, we cannot draw mechanistic conclusions based on this observation alone.

For photothermal-based mechanisms, this linear dependence implies that the temperature increase on the Au nanostructure also scales linearly with the incident laser power[5]. This suggests that the heat dissipation rate remains constant across different absolute heating powers and is unaffected by the absolute temperature gradient between the Au nanostructure and the environment.

To facilitate quantitative comparison across different measurements, all photovoltage results are normalized to the incident laser power, displayed in the unit of "µV/mW" (µV of photovoltage per mW of incident laser power). We did not determine the incident laser power density because it requires a precise measurement of the laser beam spot size.

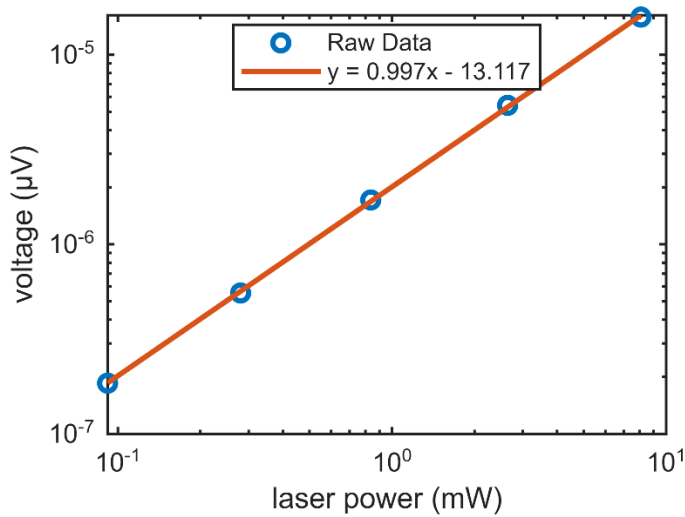

**Figure S2.** Incident laser power dependent photo-voltage, plotted in a log-log scale. The slope is very close to 1, indicating a linear relationship.

## 3. Gaussian fittings of the horizontal slices in the photovoltage map

For each horizontal slice in the photovoltage map, the measured photovoltage increases and then decreases as the focused Gaussian laser spot scans across the nanowire. This position-dependent photovoltage magnitude likely reflects the varying degrees of overlap between the laser spot and the nanowire, reaching a maximum at the position with the best overlap. Due to the 2D Gaussian intensity profile of the focused laser beam, this trend may be quantitatively represented as a 1D Gaussian intensity distribution as a first order approximation. To test this hypothesis, we fitted every horizontal slice in the photovoltage map to a 1D Gaussian intensity distribution, $V(x) = Ae^{-2\left(\frac{x-x_0}{\sigma}\right)^2}$, where $A$ is a scaling constant, $x$ is the horizontal position, $x_0$ is the center horizontal position, $\sigma$ is the $1/e^2$ radius of the intensity distribution. We then extracted the fitted $1/e^2$ radius $\sigma$ as well as the goodness-of-fit $R^2$ value from each fit (an example of the fitting in Figure S3e). Finally, we summarized the result in Figure S3c, d in correspondence with the 2D photovoltage map in Figure S3b.

We found that the extracted $\sigma$ in the "off-disks" regions is around 2 μm and agrees with the estimated $1/e^2$ radius of the focused laser beam in this study (marked as the dashed line in Figure S3c). However, the extracted $\sigma$ in the "in-disk" regions is consistently larger than 2 μm. As mentioned in the main text, this result implies the photothermal nature of the laser excitation, where the transport of photothermal heat generated on the nanodisks to the nanowire broadens the Gaussian profile, resulting in a larger $\sigma$ value. We also note that $\sigma$ in the "in-disk" regions is larger for a larger nanodisk block, reflecting on the fact that the larger nanodisk block possesses higher absorption cross section, generating more photothermal heat.

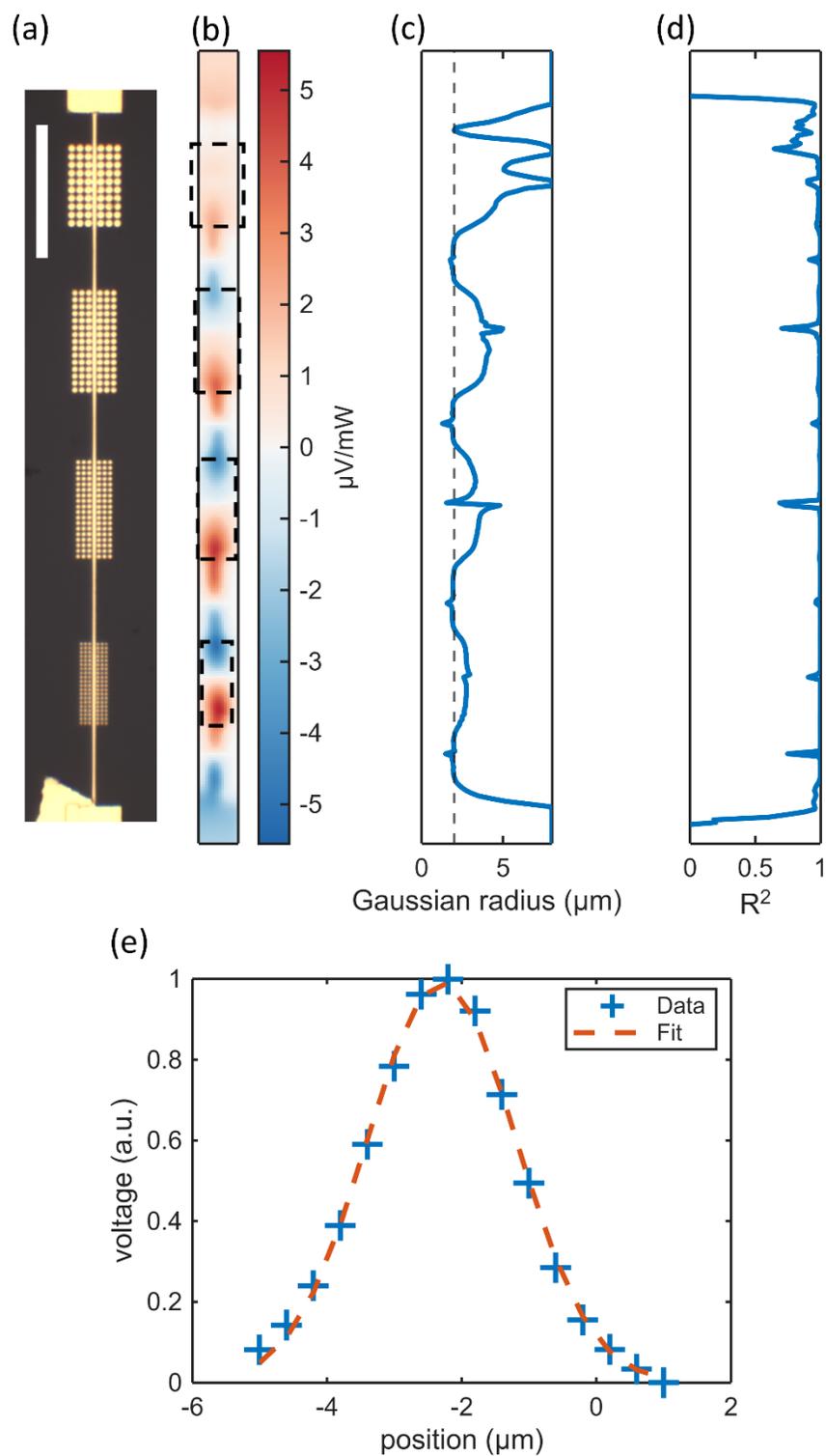

**Figure S3.** Summary of the Gaussian fittings of each horizontal slice in (b) the photovoltage map (normalized to the incident laser power) of the (a) corresponding

device. Extracted fitting parameters (c) $1/e^2$ radius $\sigma$ and (d) $R^2$ value for all horizontal slices. (e) shows an example fitting. The scale bar in (a) is 20 µm.

## 4. LDOS enhancement in the visible spectral range

We performed local photonic density of states (LDOS) simulations of the same devices as in Figure 3j (gap sizes of 120 nm, 180 nm and 220 nm, as well as a bare nanowire) using the same technique as described in Methods section, but applied to the visible spectral range of 400-800 nm (Figure S4). We observed negligible changes in LDOS enhancement on the nanowire with the presence of the nanodisks regardless of the gap size, in contrast to the significant differences in the infrared spectral range (Figure 3j). This behavior may be attributed qualitatively to the difference in relative length scale between the gap size and the wavelength of light.

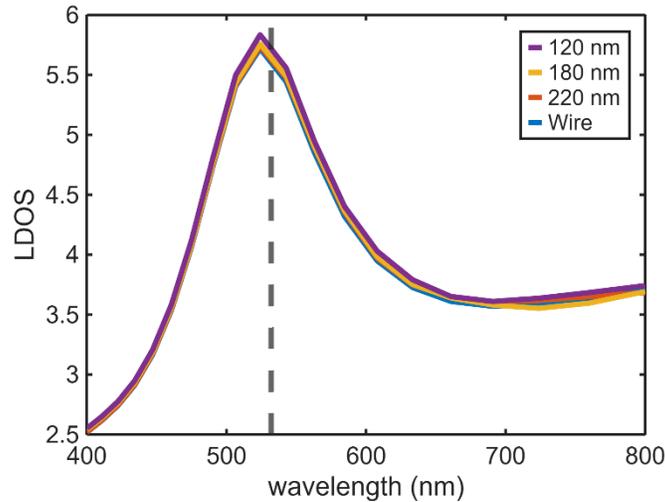

**Figure S4.** Simulated LDOS enhancement of devices with gap sizes 120 nm, 180 nm and 220 nm respectively in addition to a bare nanowire device, in the visible spectral range from 400-800 nm. Gray dashed line denotes laser wavelength of 532 nm.

## 5. Photovoltage map before and after annealing

The photovoltage map in Figure S5 corresponds to Figure 4 in the main manuscript. Here, we adjusted the color scale to show the full dynamic range of the photovoltage signal in each map, in order to better illustrate the spatial distribution of the signal. Upon closer inspection, although the magnitude of the signal is much smaller (approximately ~10%) than after 200 °C annealing (Figure S5c), the spatial distribution before annealing (Figure S5a) and after 100 °C annealing (Figure S5b) still appears roughly spatially correlated with the interfaces, exhibiting the same polarity of voltage. However, the exact spatial correlation in Figure S5a and Figure S5b is much noisier. For example, hotspots are absent at some interfaces or are shifted away from the exact interface locations. Nonetheless, these observations suggest that photonically-induced photovoltage effect was present before annealing, but hindered by the low quality factor (Q-factor) of the plasmon resonance due to abundant grain boundaries in as-deposited metal. This further corroborates with our hypothesis that annealing increased the photonic quality of the nanostructure.

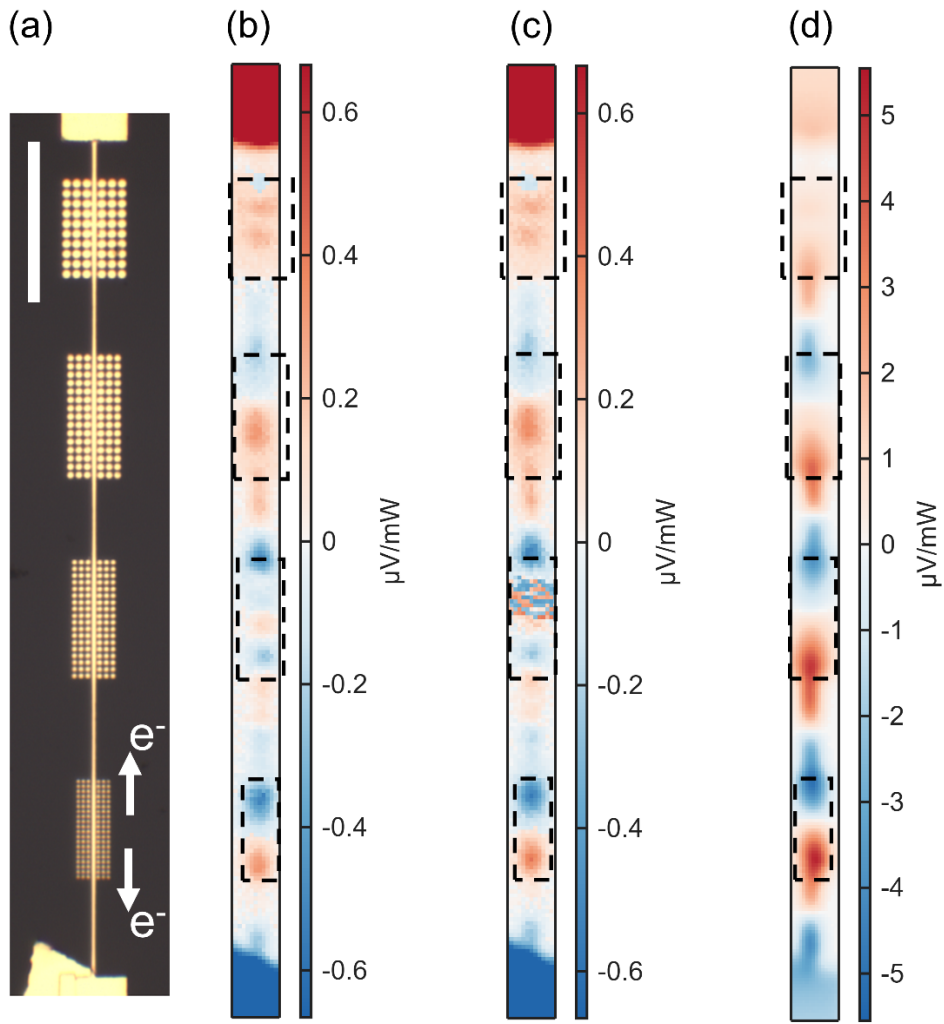

**Figure S5.** The same photovoltage maps as in Figure 4 of the main manuscript, but with adjusted color scale to show the full dynamic range. (a) Optical bright-field image of the device, where the white arrows denote the direction of electron flow as indicated by the measured photovoltage signal. The scale bar is 20 µm. (b) Photovoltage map before hot plate annealing. (b) Photovoltage map after 100 °C hot plate annealing. (c) Photovoltage map after 200 °C hot plate annealing. In (b-d), the photovoltage was normalized to the incident laser power. The black dashed lines denote the corresponding locations of the nanodisk blocks.

## 6. Photovoltage map on a bare nanowire device

The photovoltage distribution along a bare nanowire device in the region away from the terminal pads appears "random" (Figure S6b). The magnitude of the photovoltage did not significantly increase after annealing at 200 °C (Figure S6c).

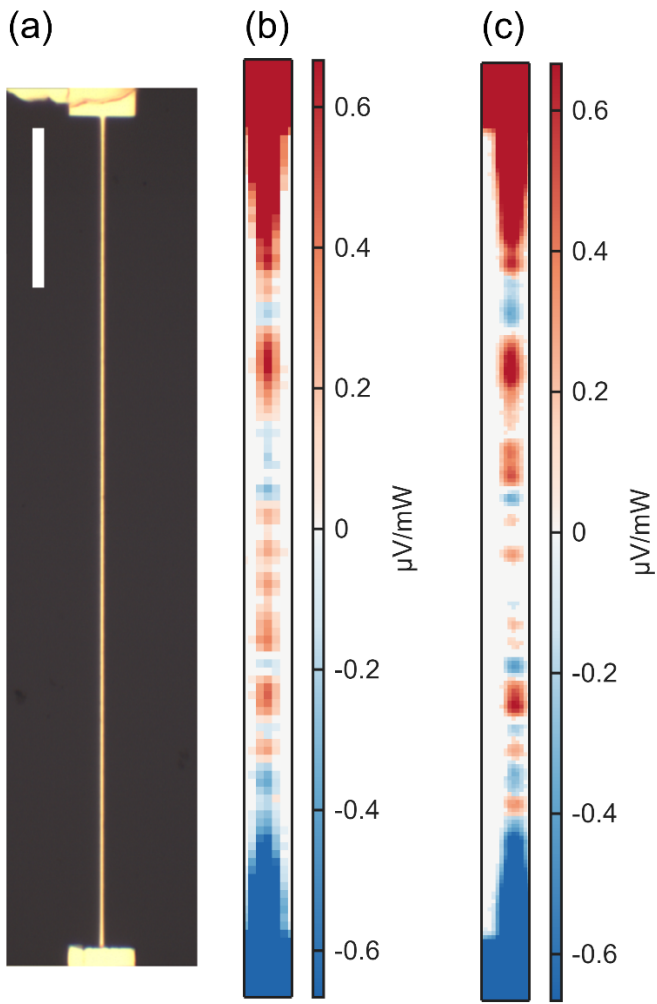

**Figure S6.** The photovoltage maps of a bare nanowire device. (a) Optical bright-field image of the device. The scale bar is 20 μm. (b) Photovoltage map before hot plate annealing. (b) Photovoltage map after 200 °C hot plate annealing. In (b-c), the photovoltage was normalized to the incident laser power. The color scale was adjusted to highlight the signal on the nanowire.

## 7. Optical images before and after annealing

We compared the optical images (bright-field and dark-field) of the device discussed in the main manuscript before and after 200 °C annealing, shown in Figure S7. There are no observable differences from either the bright-field or the dark-field optical images before and after annealing, suggesting that no major geometrical transformations occurred during annealing.

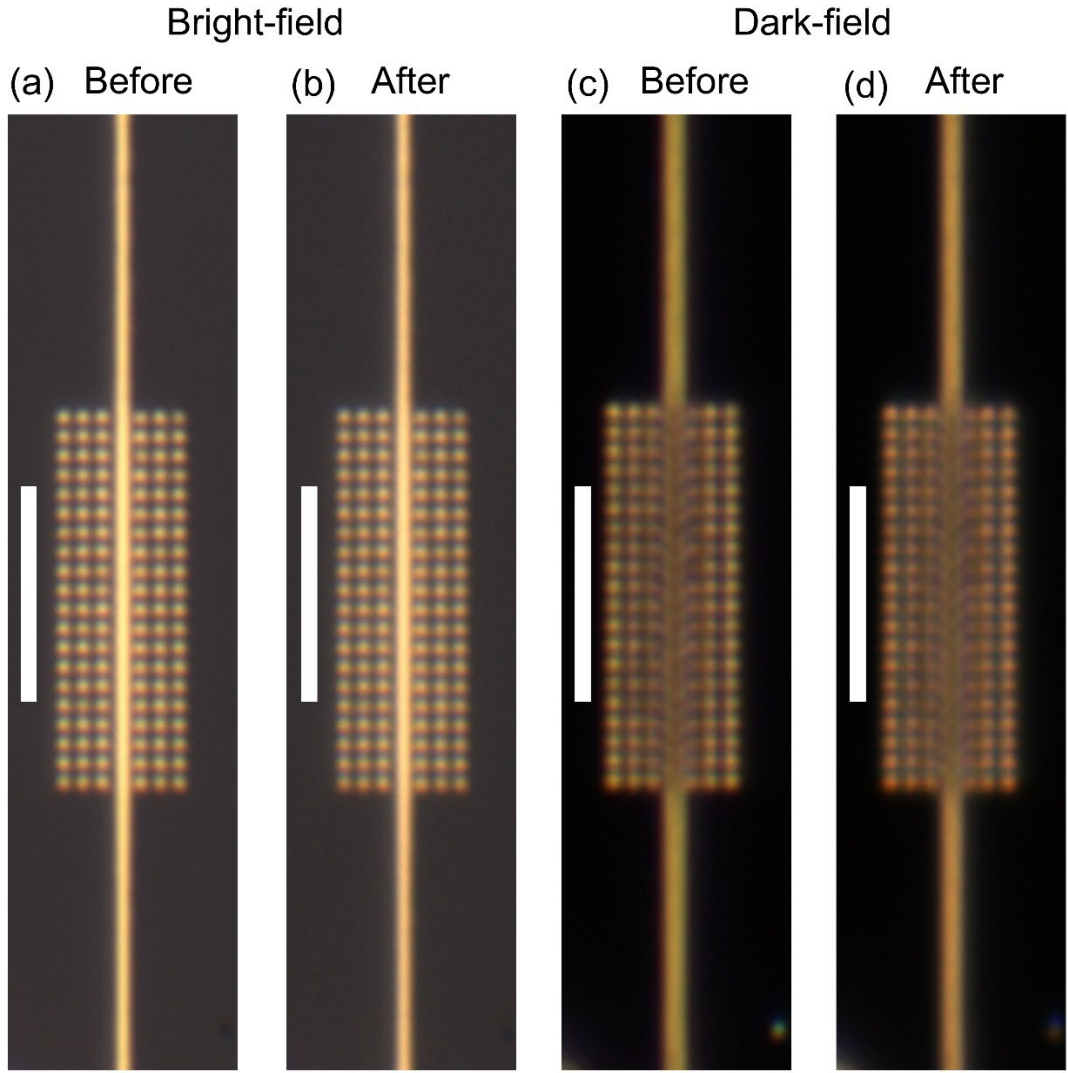

**Figure S7.** Optical images of the device (around the smallest nanodisk block, 600 nm periodicity) before and after 200 °C annealing. (a)(b) are bright-field and (c)(d) are dark-field. (a)(c) are before annealing and (b)(d) are after 200 °C annealing. All scale bars are 7 μm.

## 8. Possible intrinsic geometrical variations of the nanowire

As discussed in the main manuscript, the nanowire may exhibit slight width non-uniformities. In particular, systematic nanowire width variations may arise from the "proximity effect"[6] during electron-beam lithography, which broadens the exposed area beyond the intended pattern adjacent to other structures. This effect is mainly caused by high-energy electrons scattering within the resist and the substrate, altering the trajectories. Forward-scattered electrons propagate into nearby regions, while back-scattered electrons can travel several micrometers laterally[6] and expose resist entirely outside the directly addressed regions. In the actual nanofabrication, both the nanowire and the nanodisks were exposed in the same resist layer at the same time. The interaction length of several micrometers for the back-scattered electrons means that segments of the nanowire adjacent to the nanodisks could receive additional electron beam dose from the exposure of these nanodisks. Consequently, a larger area receives enough dose to be dissolved by the developing agent, yielding a wider Au nanowire in these regions.

Scanning electron microscope (SEM) images of the actual device reveal that the nanowire is slightly wider in the "in-disk" regions compared to the "off-disk" regions (Figure S8). We note that the nanowire broadening is most pronounced around the largest nanodisk block (Figure S8a-c), whereas the broadening is minimal around the smallest nanodisk block (Figure S8d-f). We estimated the broadening around the largest nanodisk block to be approximately 10% of the nanowire width. This contrast may be due to different nanodisk coverage in these nanodisk blocks, leading to varying proximity effects. ~55% of the area is occupied by the nanodisks in the largest block, whereas ~35% of the area is occupied by the nanodisks in the smallest block. Therefore, we focused on analyzing the photovoltage response around the smallest block, where the SEM images show negligible width variation.

We performed a rough quantitative estimate of the possible Seebeck coefficient gradient produced by a 10% nanowire width variation (assuming worse case). Our estimate is based on previously reported thermoelectric measurements of width-dependent Seebeck coefficients for thin wires of various metals, mainly Pd[7]. For a 90 nm thick Pd wire from 100 nm to 3 µm wide, the difference in the Seebeck coefficient ($\Delta S$) of this wire to the bulk value of -9 µV/K roughly follows a linear relationship to the wire width in log-log space. $\Delta S$ between a 400 nm wide wire and a 440 nm (+10%) wide wire is estimated to be approximately 0.04 µV/K, which is only 0.4% of the bulk value.

Another study reported that different metals exhibit different levels of width-induced $\Delta S$[8]. The effect on Ta is about 4.5× larger on Pd. Assuming Au exhibits $\Delta S$ as large as Ta, the Seebeck coefficient gradient would remain below 2%, which is at least an order of magnitude less than that required by the experimentally observed photovoltage.

Therefore, width variation alone cannot account for the experimentally observed spatially correlated photovoltage.

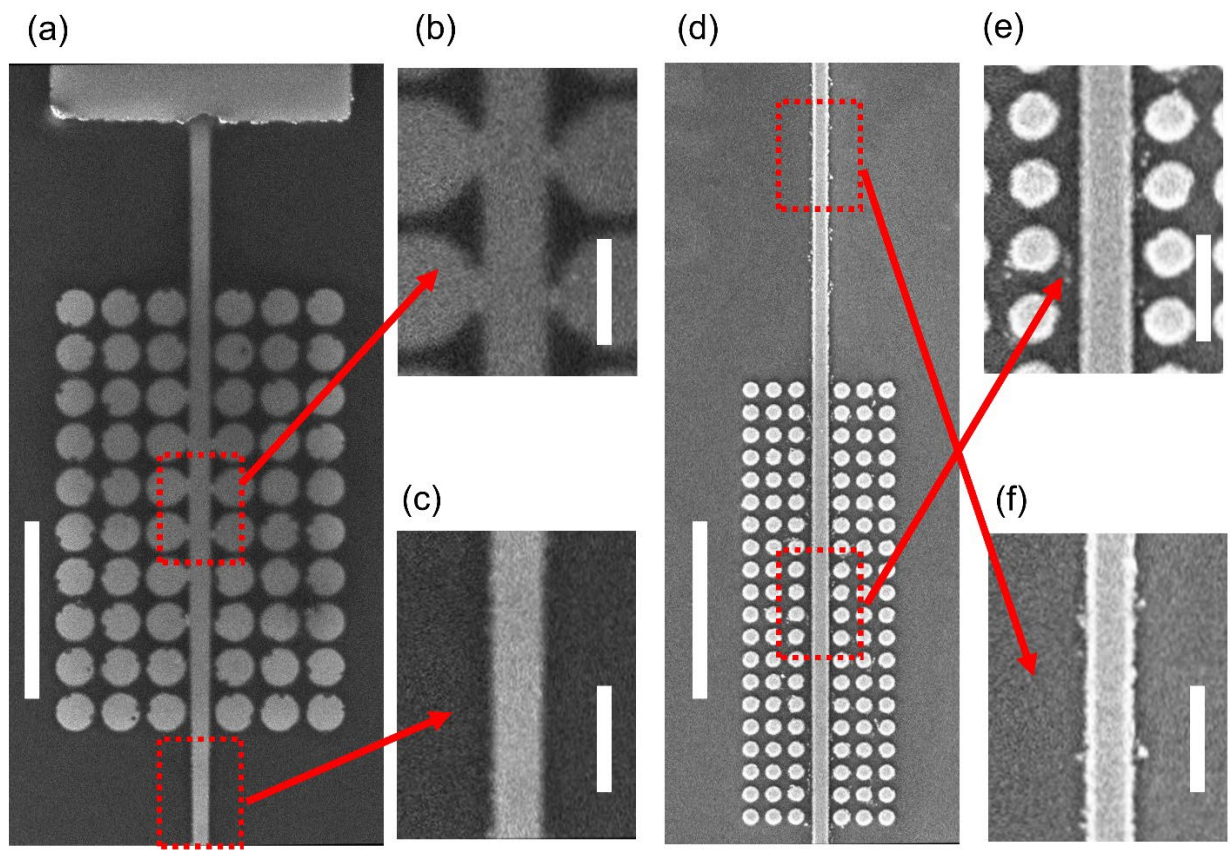

**Figure S8.** Demonstration of the nanowire width non-uniformities observed in the SEM images. (a-c) Images around the largest nanodisk block. (d-f) Images around the smallest block. Scale bars in (a)(d) are 5 µm. Scale bars in (b)(c)(e)(f) are 1 µm.